\documentclass[runningheads,a4paper]{llncs}

\usepackage{amssymb}

\usepackage[final]{graphicx}
\usepackage{url}
\newcommand{\keywords}[1]{\par\addvspace\baselineskip\noindent\keywordname\enspace\ignorespaces#1}

\usepackage{amsmath,amsfonts,bbm}
\usepackage{color}
\usepackage{algorithm2e}
\usepackage{algorithmic}
\usepackage{listings}

\usepackage{float}

\DeclareMathOperator*{\argmin}{arg\,min} 
\newcommand{\R}{\mathbb R}

\newcommand{\M}{\mathcal{M}}

\renewcommand{\exp}{\mathrm{Exp}}
\renewcommand{\log}{\mathrm{Log}}
\usepackage{listings}
\usepackage[usenames,dvipsnames,svgnames,x11names]{xcolor} 
\lstdefinestyle{mystyle}{
    language=Python,
    commentstyle=\color{DarkSlateGray4}\itshape,%\color{codegreen},
    keywordstyle=\color{Green},
    morecomment=[s][\color{Red3}\itshape\ttfamily]{"""}{"""},
    basicstyle=\small\ttfamily, 
    breakatwhitespace=false,         
    breaklines=true,                 
    captionpos=b,
    columns=fullflexible,                    
    showspaces=false,                
    showstringspaces=false,
    showtabs=false,                  
    tabsize=2,
    moredelim=**[is][\color{blue}]{@}{@} 
}
\newlength\myindent
\begin{document}
\lstset{language=Python,style=mystyle}

\mainmatter

\title{Computational Anatomy in Theano}

%\author{***}
\author{Line K\"uhnel and Stefan Sommer}
\institute{Department of Computer Science, University of Copenhagen\\
\url{kuhnel@di.ku.dk},\ \url{sommer@di.ku.dk}}

\maketitle
\vspace{-0.5cm}
\begin{abstract}
To model deformation of anatomical shapes, non-linear statistics are required to take into account the non-linear structure of the data space. Computer implementations of non-linear statistics and differential geometry algorithms often lead to long and complex code sequences. The aim of the paper is to show how the Theano framework can be used for simple and concise implementation of complex differential geometry algorithms while being able to handle complex and high-dimensional data structures. We show how the Theano framework meets both of these requirements. The framework provides a symbolic language that allows mathematical equations to be directly translated into Theano code, and it is able to perform both fast CPU and GPU computations on high-dimensional data. We show how different concepts from non-linear statistics and differential geometry can be implemented in Theano, and give examples of the implemented theory visualized on landmark representations of Corpus Callosum shapes.
\vspace{-0.2cm}

\end{abstract}
%\vspace{-0.9cm}
\section{Introduction}
\label{sec:in}
\vspace{-0.2cm}
Euclidean statistical methods can generally not be used to analyse anatomical shapes because of the non-linearity of shape data spaces. Taking into account non-linearity and curvature of the data space in statistical analysis often requires implementation of concepts from differential geometry.

 Numerical implementation of even simple concepts in differential geometry is often a complex task requiring manual implementation of long and complicated expressions involving high-order derivatives. We propose to use the Theano framework in Python to make implementation of differential geometry and non-linear statistics algorithms a simpler task. One of the main advantages of Theano is that it can perform symbolic calculations and take symbolic derivatives of even complex constructs such as symbolic integrators. As a consequence, mathematical equations can almost directly be translated into Theano code. For more information on the Theano framework, see~\cite{theano}.

\begin{figure}[H]
    \begin{center}
        \includegraphics[scale=0.4, trim = 20 20 40 30,clip]{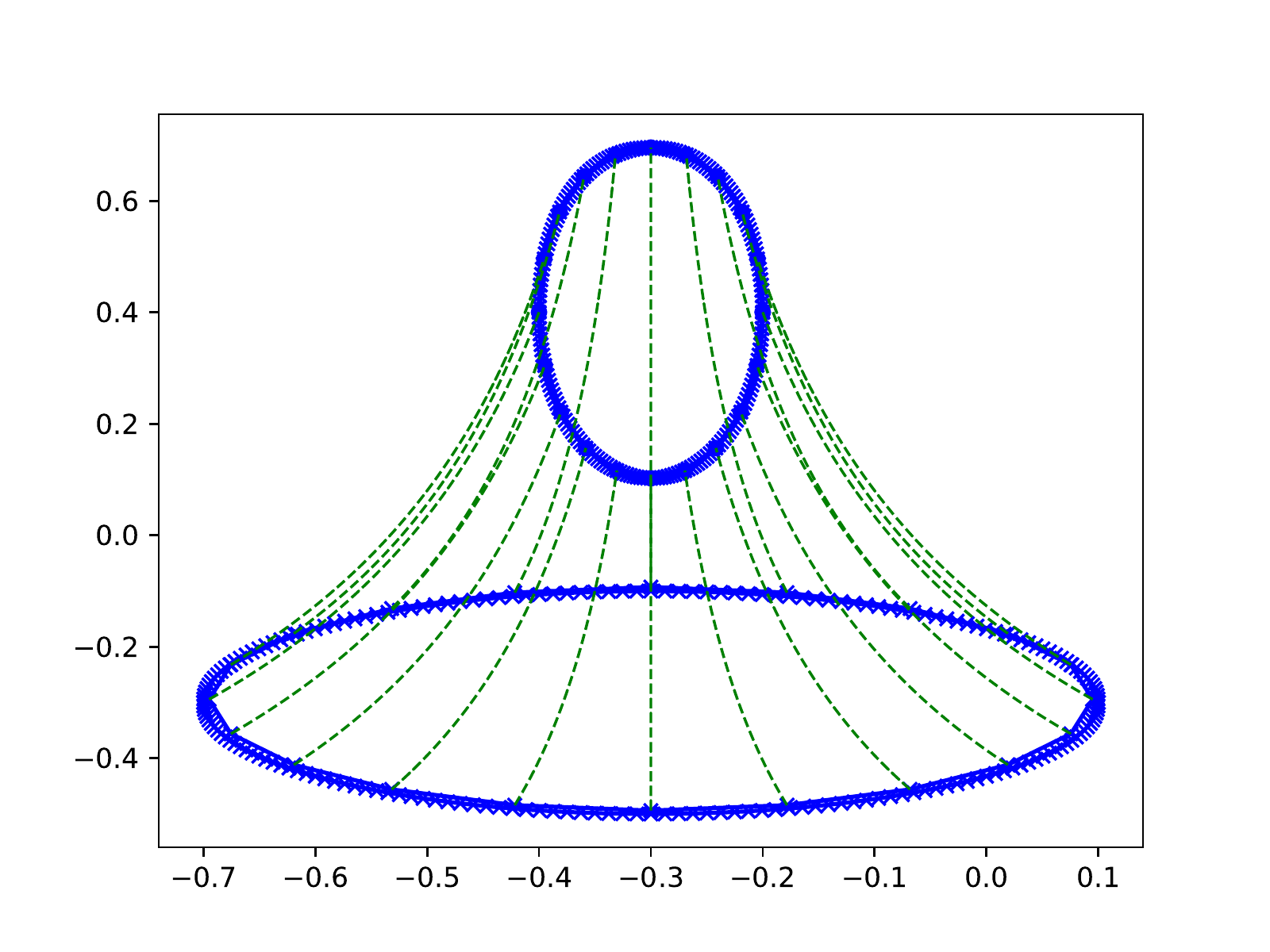}
        \caption{Matching of $20000$ landmarks on two ellipsoids. Only the matching curve for $20$ landmarks have been plotted to make the plot interpretable. The GPU computation is automatic in Theano and no explicit GPU code is used for the implementation.}
        \label{fig:match}
    \end{center}
\end{figure}

 Even though Theano make use of symbolic calculations, it is still able to perform fast computations on high-dimensional data. A main reason why Theano can handle complicated data is the opportunity to use both CPU and GPU for calculations. As an example, Fig. \ref{fig:match} shows matching of $20000$ landmarks on two different ellipsoids performed on a $40000$-dimensional landmark manifold. The matching code was implemented symbolically using no explicit GPU code.

The paper will discuss multiple concepts in differential geometry and non-linear statistics relevant to computational anatomy and provide corresponding examples of Theano implementations. We start by considering simple theoretical concepts and then move to more complex constructions from sub-Riemannian geometry on fiber bundles. Examples of the implemented theory will be shown for landmark representations of Corpus Callosum shapes using the Riemannian manifold structure on the landmark space defined in the Large Deformation Diffeomorphic Metric Mapping (LDDMM) framework.
\vspace{-0.7cm} 
\vspace{-0.2cm}

The presented Theano code is available in the Theano Geometry repository \textcolor{blue}{\url{http://bitbucket.org/stefansommer/theanogeometry}} that includes Theano implementations of additional differential geometry, Lie group, and non-linear statistics algorithms. The described implementations are not specific to the LDDMM landmark manifold used for examples here. The code is completely general and can be directly applied to analysis of data modelled in spaces with different non-linear structures. For more examples of Theano implementation of algorithms directly targeting landmark dynamics, see~\cite{arnaudon_stochastic_2016,arnaudon_geometric_2017}.

The paper is structured as follows. Section \ref{sec:LDDMM} gives a short introduction to the LDDMM manifold. Section \ref{sec:geo} concerns Theano implementation of geodesics as solution to Hamilton's equations. In Section \ref{sec:chris}, we use Christoffel symbols to define and implement parallel transport of tangent vectors. In Section \ref{sec:FMean}, the Frech\'et mean algorithm is considered, while stochastics, Brownian motions, and normal distributions are described in Section \ref{sec:Norm}. Section \ref{sec:FMeanfm} gives an example of calculating sample mean and covariance by estimating the Frech\'et mean on the frame bundle. The paper ends with concluding remarks.

\subsection{Background}
\label{sec:LDDMM}
\vspace{-0.1cm}
The implemented theory is applied to data on a landmark manifold defined in the LDDMM framework~\cite{shapes}.
More specifically, we will exemplify the theoretical concepts with landmark representations of Corpus Callosum (CC) shapes. 

\begin{figure}[H]
    \begin{center}
    \begin{minipage}{0.5\textwidth}
        \centering
        \includegraphics[scale=0.4, trim = 20 20 40 30,clip]{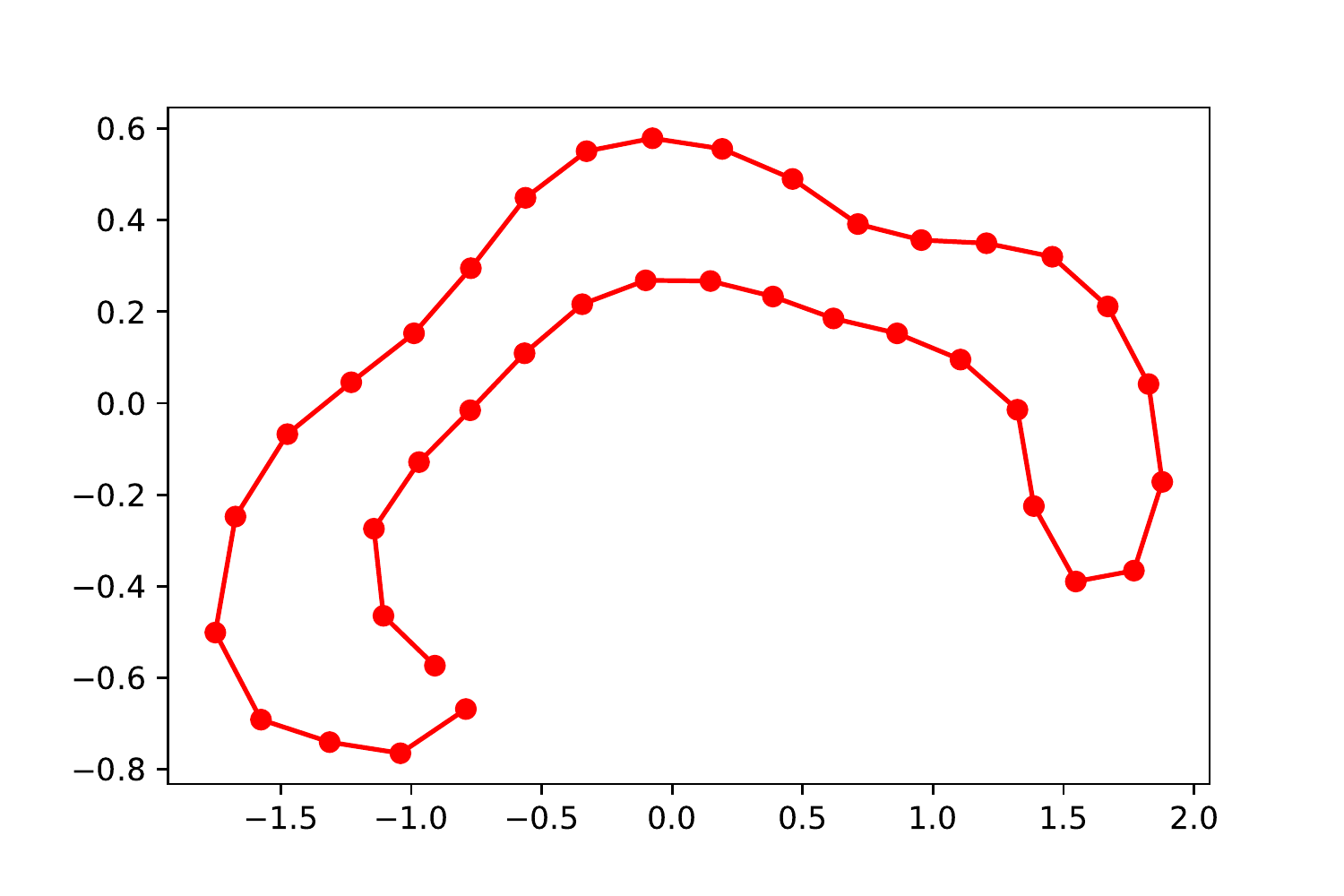}
    \end{minipage}%
    \begin{minipage}{0.5\textwidth}
        \centering
        \includegraphics[scale=0.4,trim = 20 20 40 30,clip]{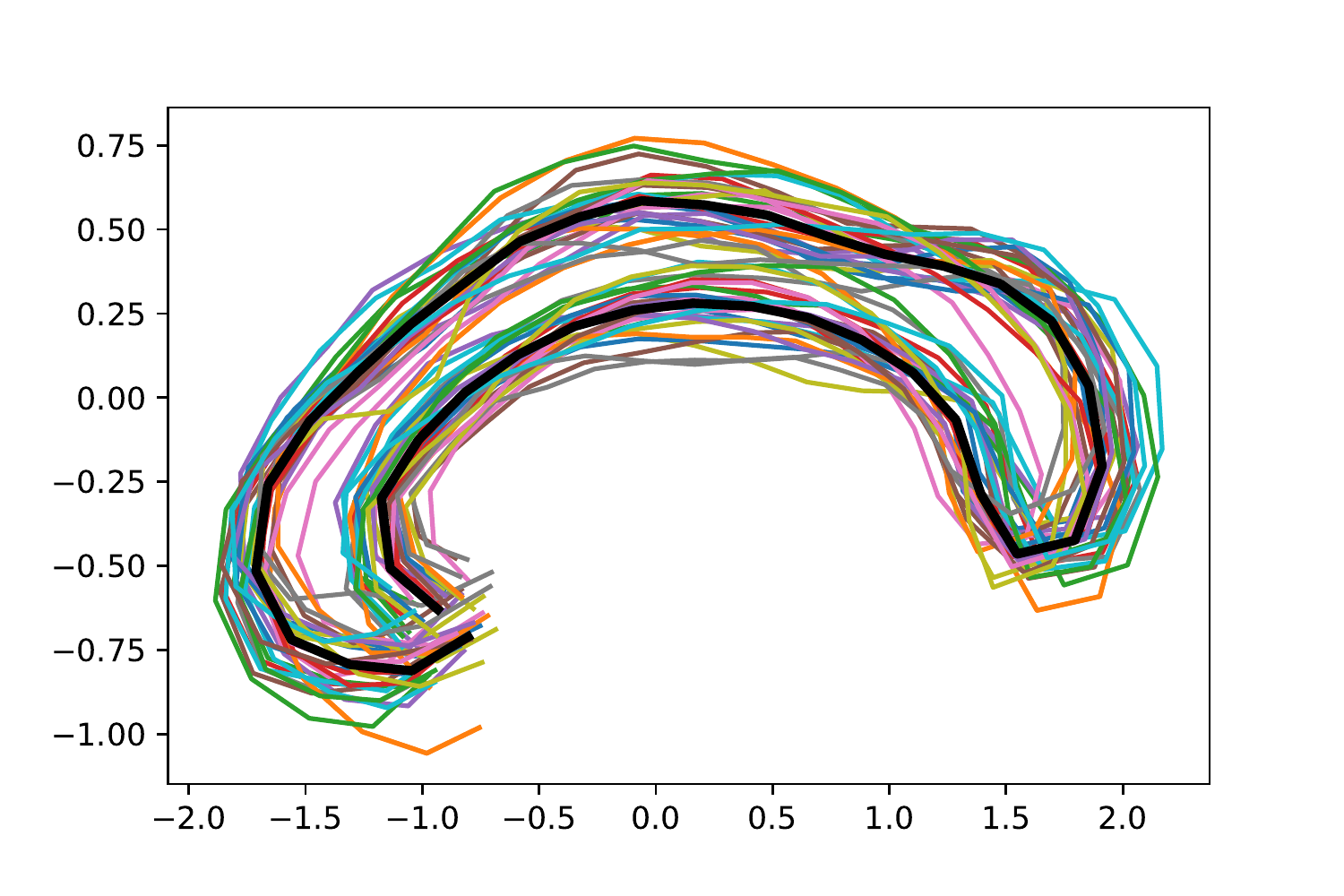}
    \end{minipage}
    \caption{(left) An example of a point in $\M$. (right) A subset of the data considered in the examples of this paper. The black curve represents the mean CC of the data.}
    \label{fig:data}
    \end{center}
\end{figure}

Consider a landmark manifold, $\M$, with elements $q = (x_1^1,x_1^2,\ldots,x_n^1,x_n^2)$ as illustrated in Fig. \ref{fig:data}. In the LDDMM framework, deformation of shapes are modelled as a flow of diffeomorphisms. Let $V$ denote a Reproducing Kernel Hilbert Space (RKHS) of vector fields and let $K\colon V\times V\to\R$ be the reproducing kernel, i.e. a vector field $v\in V$ satisfies $v(q) = \langle K_q,v\rangle_V$ for all $q\in\M$ with $K_q = K(.,q)$. Deformation of shapes in $\M$ are then modelled by flows $\varphi_t$ of diffeomorphisms acting on the landmarks. The flow solves the ordinary differential equation $\partial_t\varphi_t = v(t)\circ\varphi_t$, for $v\in V$. 
With suitable conditions on $K$, the norm on $V$ defines a right-invariant metric on the diffeomorphism group that descends to a Riemannian structure on $\M$. The induced cometric $g^*_q\colon T^*_q\M\times T^*_q\M\to\R$ takes the form
\begin{equation}
    g^*_q(\nu,\xi) = \sum_{i,j=1}^n \nu_i K(\boldsymbol{x}_i,\boldsymbol{x}_j)\xi_j,
\label{eq:metric}
\end{equation}
where $\boldsymbol{x}^i = (x_i^1,x_i^2)$ for $i\in\{1,\ldots,n\}$. The coordinate matrix of the cometric is $g^{ij}=K(\boldsymbol{x}_i,\boldsymbol{x}_j)$ which results in the metric $g$ having coordinates $g_{ij} = K^{-1}(\boldsymbol{x}_i,\boldsymbol{x}_j)$.

In the examples, we use $39$ landmarks representing the CC shape outlines, and the kernel used is a Gaussian kernel defined by $K(\boldsymbol{x}_i,\boldsymbol{x}_j) = \exp\left(-\frac{\|\boldsymbol{x}_i-\boldsymbol{x}_j\|^2}{2\sigma^2}\right)$ with variance parameter $\sigma$ set to the average distance between landmarks in the CC data. Samples of CC outlines are shown in the right plot of Fig. \ref{fig:data}.
\vspace{-0.4cm}

%\vspace{-1cm}
\section{Geodesics}
\label{sec:geo}
\vspace{-0.2cm}
 Geodesics on $\M$ can be obtained as the solution to Hamilton's equations used in Hamiltonian mechanics to describe the change in position and momentum of a particle in a physical system.
Let $(U,\varphi)$ be a chart on $\M$ and assume $(\M,g)$ is a Riemannian manifold. The Hamiltonian $H$ describes the total amount of energy in the physical system. From the cometric $g^*$, the Hamiltonian can be defined as $H(q,p) = \frac{1}{2}p^Tg^*_qp$, where $g^*_q=(g^{ij})$ is the component matrix of $g^*$ at $q$. Hamilton's equations are given as the system of ordinary differential equations
\begin{align*}
    dq_t = \nabla_p H(q,p), \quad dp_t = -\nabla_q H(q,p).
\end{align*}
Using the symbolic derivative feature of Theano, the system of ODE's can be represented and discretely integrated with the following code snippet:
\begin{lstlisting}
"""
Hamiltonian function and equations
"""
# Hamiltonian function:
H = lambda q,p: 0.5*T.dot(p,T.dot(gMsharp(q),p))
# Hamiltonian equations:
dq = lambda q,p: T.grad(H(q,p),p)
dp = lambda q,p: -T.grad(H(q,p),q)

def @ode_Ham@(t,x):
    dqt = dq(x[0],x[1])
    dpt = dp(x[0],x[1])
    return T.stack((dqt,dpt))
# Geodesic:
Exp = lambda q,v: integrate(ode_Ham,T.stack((q,gMflat(v))))
\end{lstlisting}
where \lstinline!gMflat! is the $\flat$ map turning tangent vectors in $T\M$ to elements in $T^\star\M$. \lstinline!integrate! denotes a function that integrates the ODE by finite time discretization. For the examples considered here, we use a simple Euler integration method. Higher-order integrators are available in the implemented repository mentioned in Section \ref{sec:in}. A great advantage of Theano is that such integrators can be implemented symbolically as done below using a symbolic \lstinline!for!-loop specified with \lstinline!theano.scan!. The actual numerical scheme is only available after asking Theano to compile the function.
\begin{lstlisting}
"""
Numerical Integration Method
"""
def @integrator@(ode_f):
    def @euler@(*y):
        t = y[-2]
        x = y[-1]
        return (t+dt,x+dt*ode_f(*y))
    return euler

def @integrate@(ode,x):
    (cout, updates) = theano.scan(fn=integrator(ode),
            outputs_info=[x],sequences=[*y], n_steps=n_steps)
    return cout
\end{lstlisting}
In the above, \lstinline!integrator! specifies the chosen integration method, in this example the Euler method. As the \lstinline!integrate! function is a symbolic Theano function, symbolic derivatives can be obtained for the integrator, allowing e.g. gradient based optimization for the initial conditions of the ODE.

\begin{figure}
    \begin{center}
    \begin{minipage}{0.5\textwidth}
        \centering
        \includegraphics[scale=0.4, trim = 20 20 40 30,clip]{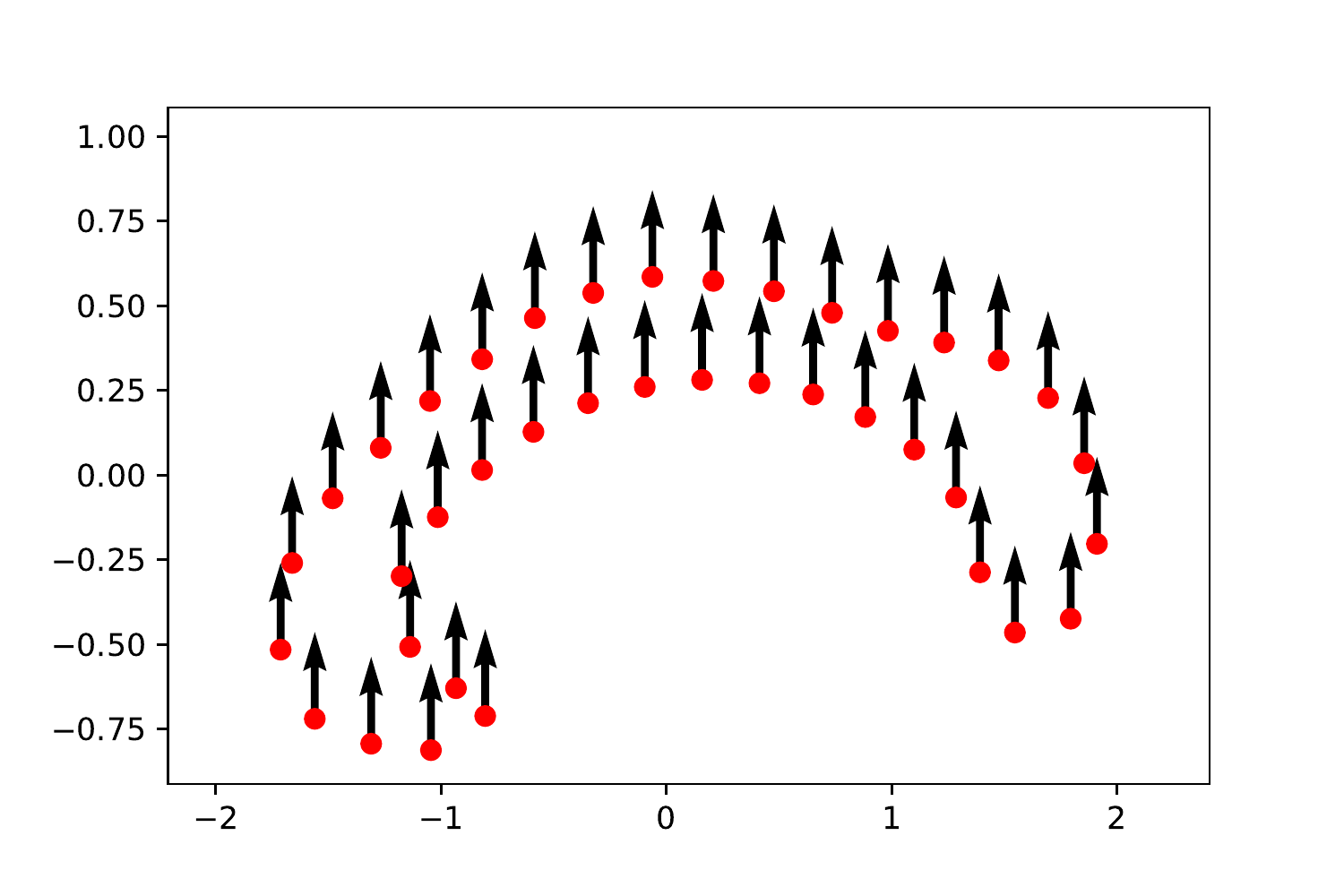}
    \end{minipage}%
    \begin{minipage}{0.5\textwidth}
        \centering
        \includegraphics[scale=0.4,trim = 20 20 40 30,clip]{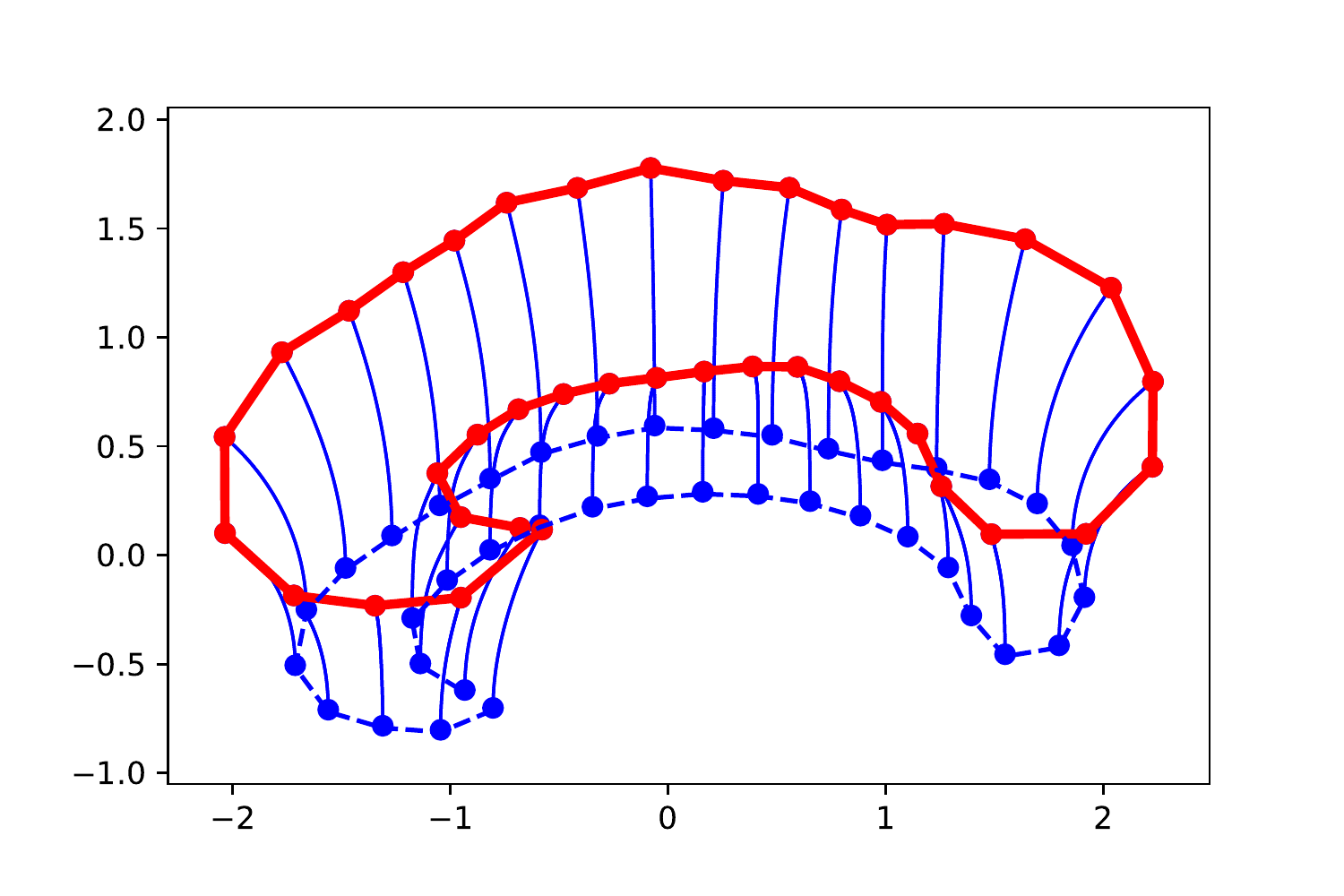}
    \end{minipage}
    \caption{(left) The initial point and tangent vector for the geodesic. (right) A geodesic obtained as solution to Hamilton's equations.}
    \label{fig:geo}
    \end{center}
\end{figure}

An example of a geodesic found as the solution to Hamilton's equations is visualized in the right plot of Fig. \ref{fig:geo}. The initial point $q_0\in\M$ was set to the average CC for the data shown in Fig. \ref{fig:data} and the initial tangent vector $v_0\in T_{q_0}\M$ was given as the tangent vector plotted in Fig. \ref{fig:geo}.
%\vspace{-0.1cm}

The exponential map, $\exp_x\colon T_x\M\to\M$, $x\in\M$ is defined as $\exp_x(v) = \gamma^v_1$, where $\gamma^v_t$, $t\in [0,1]$ is a geodesic with $\dot{\gamma}^v_0=v$. The inverse of the exponential map is called the logarithm map, denoted $\log$. Given two points $q_1,q_2\in\M$, the logarithm map retuns the tangent vector $v\in T_{q_1}\M$ that results in the minimal geodesic from $q_1$ to $q_2$, i.e. $v$ satisfies $\exp_{q_1}(v) = q_2$. The logarithm map can be implemented using derivative based optimization by taking a symbolic derivative of the exponential map, \lstinline!Exp!, implemented above:
\begin{lstlisting}
"""
Logarithm map
"""
# Loss function for landmarks:
loss = lambda v,q1,q2: 1./d*T.sum(T.sqr(Exp(q1,v)-q2))
dloss = lambda v,q1,q2: T.grad(loss(v,q1,q2),v)
# Logarithm map: (v0 initial guess)
Log = minimize(loss, v0, jac=dloss, args=(q1,q2))
\end{lstlisting}
The use of the derivative features provided in Theano to take symbolic derivatives of a discrete integrator makes the implementation of the logarithm map extremely simple. The actual compiled code internally in Theano corresponds to a discrete backwards integration of the adjoint of the Hamiltonian system. An example of matching shapes by the logarithm map was shown in Fig. \ref{fig:match}. Here two ellipsoids of $20000$ landmarks were matched by applying the above \lstinline!Log! function.

\section{Christoffel Symbols}
\label{sec:chris}
\vspace{-0.2cm}
We here describe how Christoffel symbols can be computed and used in the Theano framework.
A connection $\nabla$ defines links between tangent spaces on $\M$ and describes how tangent vectors for different tangent spaces relate. Let $(U,\varphi)$ denote a coordinate chart on $\M$ with basis coordinates $\partial_i$, $i=1,\ldots,d$. The connection $\nabla$ is uniquely described by its Christoffel symbols, $\Gamma_{ij}^k$, defined as $\nabla_{\partial_i} \partial_j = \Gamma_{ij}^k\partial_k$.

\begin{figure}
    \begin{center}
        \includegraphics[scale=0.5, trim = 20 20 40 30,clip]{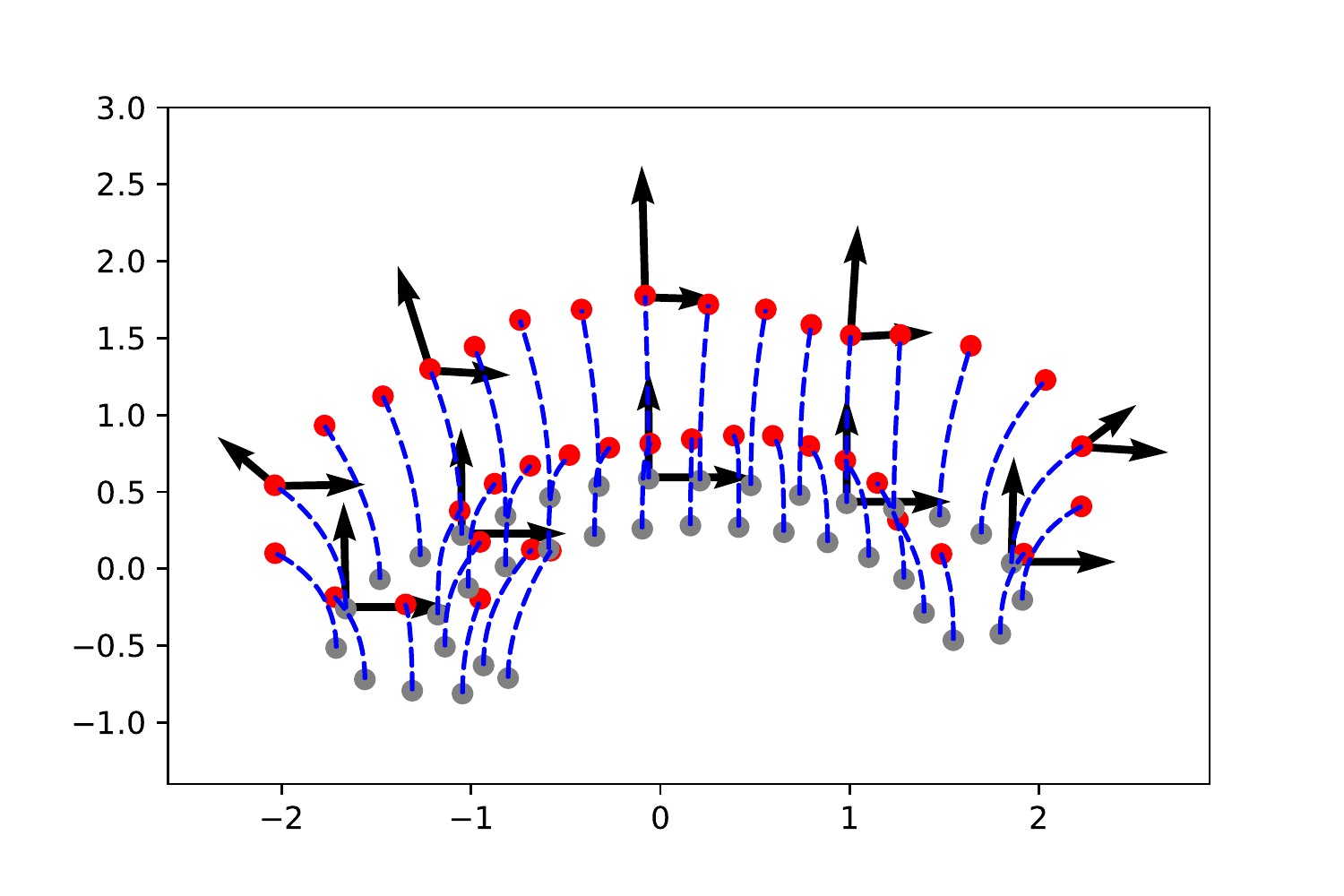}
    \caption{Example of parallel transport of basis vectors $v_1$, $v_2$ along the geodesic with initial values $q_0$, $v_2$. The parallel transported vectors are only plotted for $5$ landmarks.}
    \label{fig:par}
    \end{center}
\end{figure}

An example of a frequently used connection is the Levi-Civita connection for Riemannian manifolds. Based on the metric $g$ on $\M$, the Levi-Civita Christoffel symbols are found by
\begin{equation}
    \Gamma_{ij}^k = \frac{1}{2}g^{kl}(\partial_i g_{jl} + \partial_j g_{il} - \partial_l g_{ij}).
\label{eq:chris}
\end{equation}
The Theano implementation below of the Christoffel symbols directly translates $\eqref{eq:chris}$ into code:
\begin{lstlisting}
"""
Christoffel Symbols
"""
## Cometric:
gsharp = lambda q: T.nlinalg.matrix_inverse(g(q))    
## Derivative of metric:
Dg = lambda q: T.jacobian(g(q).flatten(),q).reshape((d,d,d))  
## Christoffel symbols:
Gamma_g = lambda q: 0.5*(T.tensordot(gsharp(q),Dg(q),axes = [1,0])\
     + T.tensordot(gsharp(q),Dg(q),axes = [1,0]).dimshuffle(0,2,1)\
     - T.tensordot(gsharp(q),Dg(q),axes = [1,2]))
\end{lstlisting}
The connection, $\nabla$, and Christoffel symbols, $\Gamma_{ij}^k$, can be used to define parallel transport of tangent vectors on $\M$. Let $\gamma\colon I\to\M$ be a curve and let $t_0\in I$. A vector field $V$ is said to be parallel along $\gamma$ if the covariant derivative of $V$ along $\gamma$ is zero, i.e. $\nabla_{\dot\gamma_t} V = 0$.
 For a tangent vector $v_0=v_0^i\partial_i\in T_{\gamma_{t_0}}\M$, there exists a unique parallel vector field $V$ along $\gamma$ s.t. $V_{t_0} = v_0$. Assume $V_t = v^i(t)\partial_i$, then the vector field $V$ is parallel along $\gamma$ if the coordinates follows the differential equation,
\begin{equation}
    \dot v^k(t) + \Gamma_{ij}^k(\gamma_t)\dot\gamma_t^i v^j(t) = 0,
\end{equation}
with initial values $v^i(0) = v_0^i$. In Theano code the ODE can be written as,
\vspace{-0.15cm}
\begin{lstlisting}
"""
Parallel transport
"""
def @ode_partrans@(gamma,dgamma,t,x): 
    dpt = - T.tensordot(T.tensordot(dgamma, Gamma_gM(gamma), 
                                    axes = [0,1]),x, axes = [1,0])
    return dpt

pt = lambda v,gamma,dgamma: integrate(ode_partrans,v,gamma,dgamma)
\end{lstlisting}
Let $q_0$ be the mean CC plotted in Fig. \ref{fig:geo} and consider $v_1,v_2\in T_{q_0}\M$ s.t. $v_1$ is the vector consisting of $39$ copies (one for each landmark) of $e_1 = (1,0)$ and $v_2$, the vector of $39$ copies of $e_2 = (0,1)$. The tangent vector $v_2$ is shown in Fig. \ref{fig:geo}. Define $\gamma$ as the geodesic calculated in Section \ref{sec:geo} with initial values $(q_0,v_2)$. The parallel transport of $v_1,v_2$ along $\gamma$ is visualized in Fig \ref{fig:par}. To make the plot easier to interpret, the parallel transported vectors are only shown for five landmarks.
\vspace{-0.2cm}

\section{Frech\'et Mean}
\label{sec:FMean}
\vspace{-0.2cm}
The Fr\'echet mean~\cite{fréchet1944intégrale} is a generalization of the Euclidean mean value to manifolds. Let $d$ be a distance map on $\M$. The Frech\'et mean set is defined as $F(x) = \argmin_{y\in\M}\mathbb{E} d(y,x)^2$. For a sample of data points $x_1,\ldots,x_n\in\M$, the empirical Fr\'echet mean is
\begin{equation}
\label{eq:FMean}
    F_{\bar{x}} = \argmin_{y\in\M} \frac{1}{n}\sum_{i=1}^n d(y,x_i)^2.
\end{equation}
Letting $d$ be the Riemannian distance function determined by the metric $g$, the distance can be formulated in terms of the logarithm map, defined in Section \ref{sec:geo}, as $d(x,y) = \|\log(x,y)\|^2$. In Theano, the Fr\'echet mean can be obtained by optimizing the function implemented below, again using symbolic derivatives. 
\begin{lstlisting}
"""
Frechet Mean
"""
def @Frechet_mean@(q,y):
    (cout,updates) = theano.scan(fn=loss, non_sequences=[v0,q],
                sequences=[y], n_steps=n_samples)
    return 1./n_samples*T.sum(cout)
dFrechet_mean = lambda q,y: T.grad(Frechet_mean(q,y),q)     
\end{lstlisting}
The variable \lstinline!v0! denotes the optimal tangent vector found with the \lstinline!Log! function in each iteration of the optimization procedure.

\begin{figure}
    \begin{center}
    \begin{minipage}{0.5\textwidth}
        \centering
        \includegraphics[scale=0.4, trim = 20 20 40 30,clip]{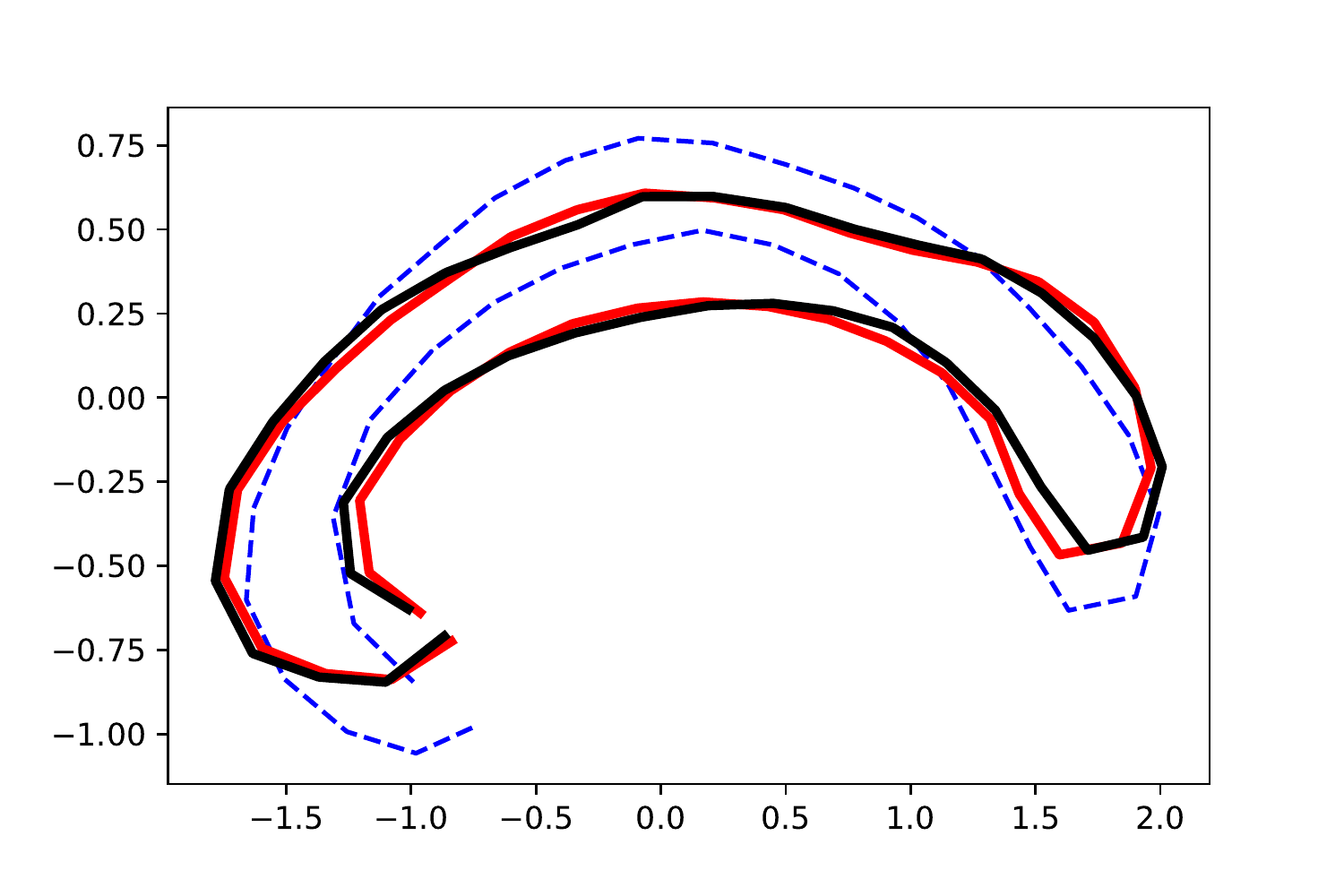}
    \end{minipage}%
    \begin{minipage}{0.5\textwidth}
        \centering
        \includegraphics[scale=0.4,trim = 20 20 40 30,clip]{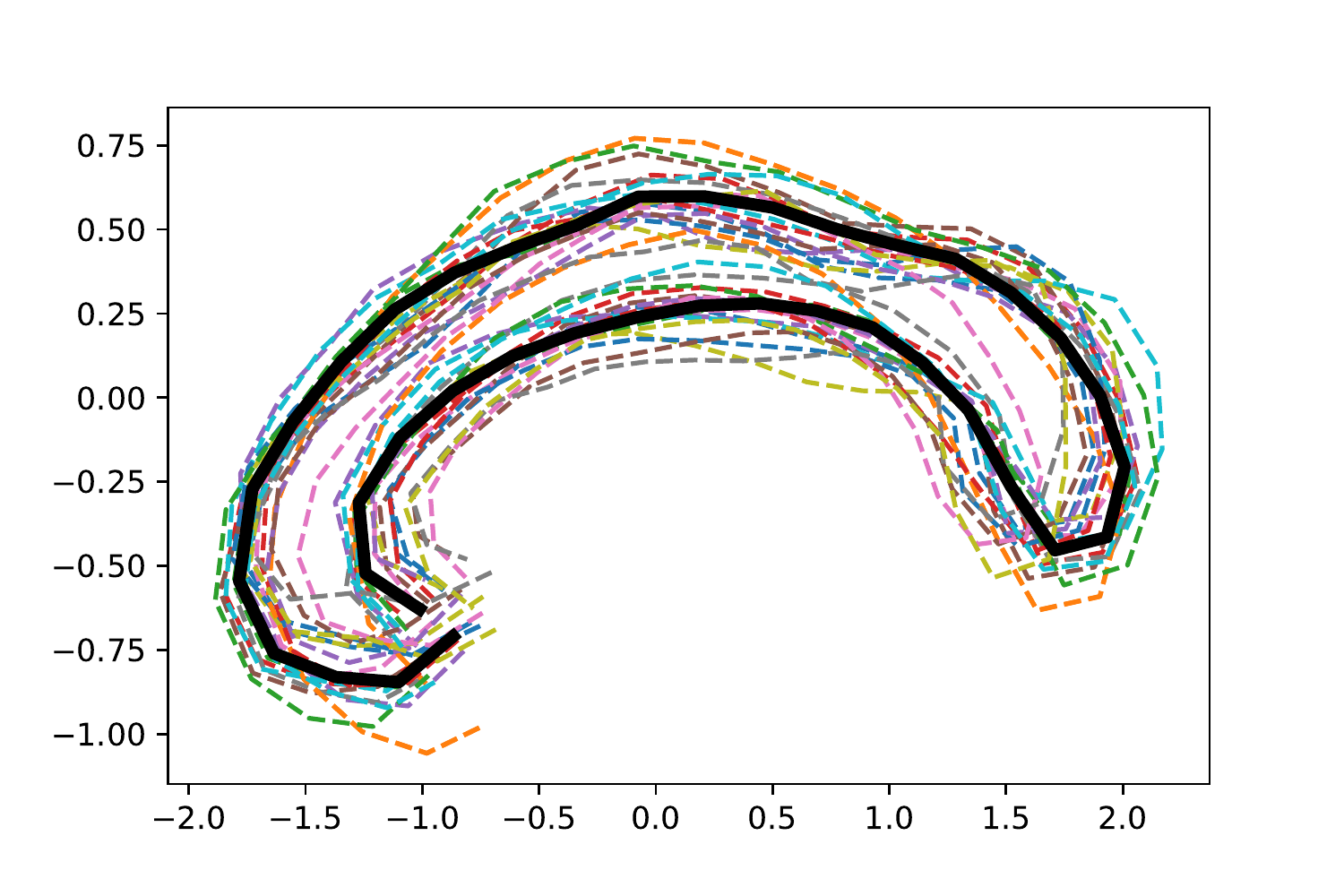}
    \end{minipage}
    \caption{(left) The estimated empirical Frech\'et mean (black), the initial value (blue) and the Euclidean mean of the $20$ samples (red). (right) Plot of the $20$ samples of CC, with the Frech\'et mean shown as the black curve.}
    \label{fig:frechet}
    \end{center}
\end{figure}

Consider a sample of $20$ observations of the CC data shown in the right plot of Fig. \ref{fig:frechet}. To calculate the empirical Fr\'echet mean on $\M$, the initial point $q_0\in\M$ was set to one of the CC observations plotted in the left plot of Fig. \ref{fig:frechet}. The result of the optimization is shown in Fig. \ref{fig:frechet} (bold outline).

So far we have shown how Theano can be used to implement simple and frequently used concepts in differential geometry. In the following sections, we will exemplify how Theano can be used for stochastic dynamics and for implementation of more complex concepts from sub-Riemannian geometry on the frame bundle of $\M$.

\section{Normal Distributions and Stochastic Development}
\label{sec:Norm}
\vspace{-0.2cm}
We here  consider Brownian motion and normal distributions on manifolds. Brownian motion on Riemannian manifolds can be constructed in several ways. Here, we consider two definitions based on stochastic development and coordinate representation of Brownian motion as an It\^o SDE. The first approach \cite{elworthy_geometric_1988} allows anisotropic generalizations of the Brownian motion \cite{sommer_modelling_2015,sommer_anisotropic} as we will use later.

Stochastic processes on $\M$ can be defined by transport of processes from $\R^m$, $m\leq d$ to $\M$ by the stochastic development map. In order to describe stochastic development of processes onto $\M$, the frame bundle has to be considered. 

The frame bundle, $F\M$, is the space of points $u=(q,\nu)$ s.t. $q\in\M$ and $\nu$ is a frame for the tangent space $T_q\M$. The tangent space of $F\M$, $TF\M$, can be split into a vertical subspace, $VF\M$, and a horizontal subspace, $HF\M$, i.e. $TF\M = VF\M\oplus HF\M$. The vertical space, $VF\M$, describes changes in the frame $\nu$, while $HF\M$ defines changes in the point $x\in\M$ when the frame $\nu$ is fixed in the sense of having zero acceleration measured by the connection. The frame bundle can be equipped with a sub-Riemannian structure by considering the distribution $HF\M$ and a corresponding degenerate cometric $g^*_{F\M}\colon TF\M^*\to HF\M$. Let $(U,\varphi)$ denote a chart on $\M$ with coordinates $(x^i)_{i=1,\ldots,d}$ and coordinate frame $\partial_i = \frac{\partial}{\partial x^i}$ for $i=1,\ldots,d$. Let $\nu_\alpha$ $\alpha=1,\ldots,d$ denote the basis vectors of the frame $\nu$. Then $(q,\nu)$ have coordinates $(q^i,\nu_\alpha^i)$ where $\nu_\alpha = \nu_\alpha^i\partial_i$ and $\nu^\alpha_i$ defines the inverse coordinates of $\nu_\alpha$. The coordinate representation of the sub-Riemannian cometric is then given as
\begin{equation}
    (g_{F\M})^{ij} = \begin{pmatrix} 
W^{-1} & -W^{-1}\Gamma^T \\
-\Gamma W^{-1} & \Gamma W^{-1}\Gamma^T
\end{pmatrix},
\end{equation}
where $W$ is the matrix with components $W_{ij} = \delta_{\alpha\beta}\nu_i^\alpha\nu_j^\beta$ and $\Gamma = (\Gamma_j^{h_\gamma})$ for $\Gamma_j^{h_\gamma}=\Gamma^h_{ji}\nu_\gamma^i$ with $\Gamma^h_{ji}$ denoting the Christoffel symbols for the connection, $\nabla$.
The sub-Riemannian structure restricts infinitesimal movements to be only along horizontal tangent vectors. Let $\pi_\nu^*\colon T_\nu\M\to H_\nu F\M$ be the lift of a tangent vector in $T\M$ to its horizontal part and let $e\in\R^d$ be given. A horizontal vector at $u=(q,\nu)$ can be defined as the horizontal lift of the tangent vector $\nu e\in T_q\M$, i.e. $H_e(u) = (\nu e)^*$. A basis for the horizontal subspace at $u\in F\M$ is then defined as $H_i = H_{e_i}(u)$, where $e_1,\ldots,e_d$ denote the canonical basis of $\R^d$.

\begin{figure}
    \begin{center}
    \begin{minipage}{0.5\textwidth}
        \centering
        \includegraphics[scale=0.4, trim = 20 20 40 30,clip]{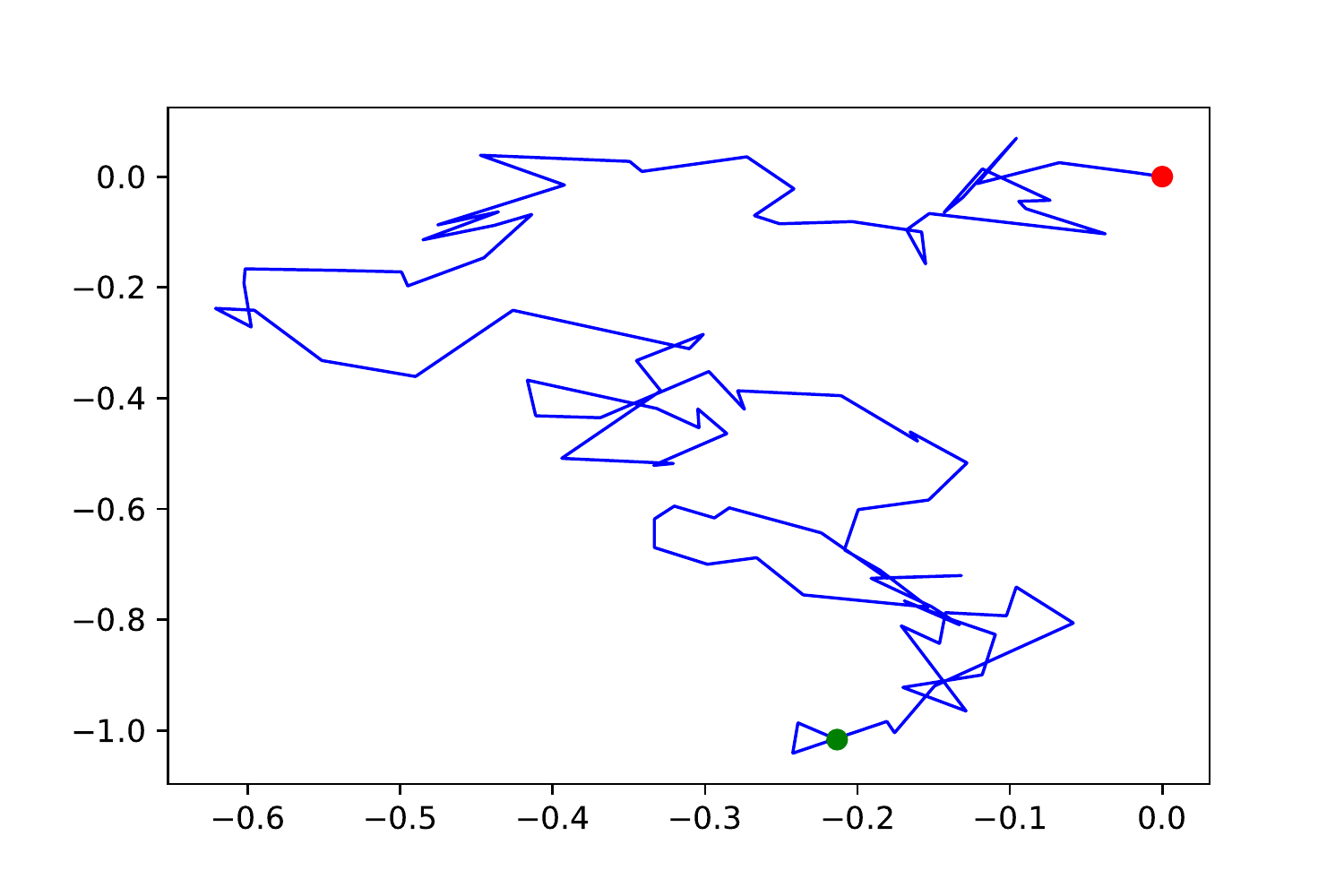}
    \end{minipage}%
    \begin{minipage}{0.5\textwidth}
        \centering
        \includegraphics[scale=0.4,trim = 20 20 40 30,clip]{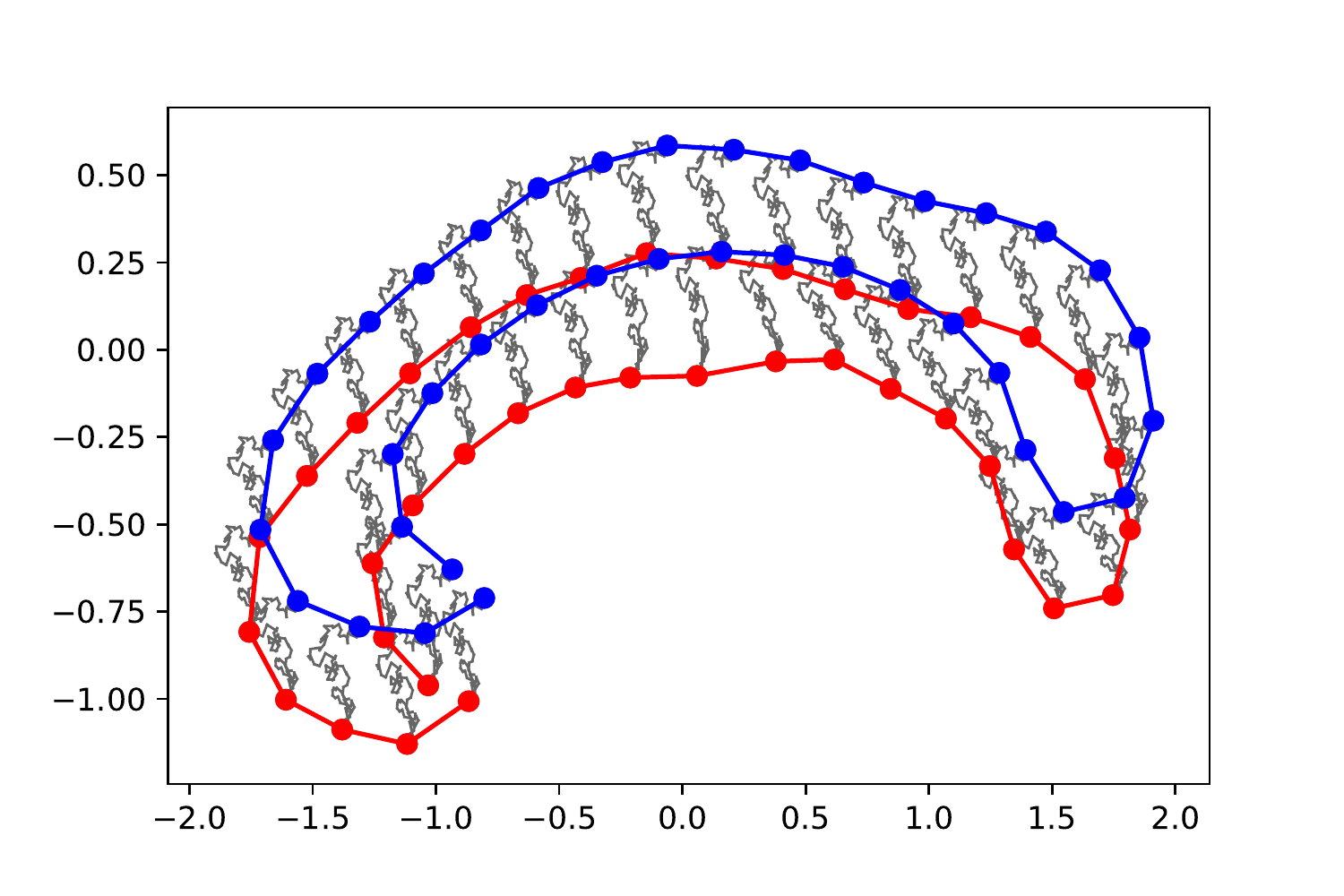}
    \end{minipage}
    \caption{(left) Stochastic process $W_t$ on $\R^2$. (right) The stochastic development of $W_t$ on $\M$. The blue points represents the initial point chosen as the mean CC. The red points visualize the endpoint of the stochastic development.}
    \label{fig:stoc}
    \end{center}
\end{figure}

Let $W_t$ denote a stochastic process on $\R^m$, $m\leq d$. A stochastic process $U_t$ on $F\M$ can be obtained by the solution to the stratonovich stochastic differential equation, $dU_t = \sum_{i=1}^m H_i(U_t)\circ dW^i_t$, with initial point $u_0\in F\M$. A stochastic process on $\M$ can then be defined as the natural projection of $U_t$ to $\M$. In Theano, the stochastic development map is implemented as
\begin{lstlisting}
"""
Stochastic Development
"""
def @sde_SD@(dWt,t,q,nu): 
    return T.tensordot(Hori(q,nu), dWt, axes = [1,0])
stoc_dev = lambda q,u,dWt: integrate_sde(sde_SD,
                       integrator_stratonovich,q,u,dWt)[1]
\end{lstlisting}
Here, \lstinline!integrate_sde! is a function performing stochastic integration of the SDE. The \lstinline!integrate_sde! is defined in a similar manner as \lstinline!integrate! described in Section \ref{sec:geo}. In Fig. \ref{fig:stoc} is given an example of stochastic development of a stochastic process $W_t$ in $\R^2$ to the landmark manifold. Notice that for $m < d$, only the first $m$ basis vectors of the basis $H_i$ is used in the stochastic development.
\vspace{-0.4cm}

Given the stochastic development map, Brownian motions on $\M$ can be defined as the projection of the stochastic development of Brownian motions in $\R^d$. Defining Brownian motions by stochastic development makes it possible to consider Brownian motions with anisotropic covariance by choosing the initial frame as not being orthonormal. 

\begin{figure}
    \begin{center}
    \begin{minipage}{0.5\textwidth}
        \centering
        \includegraphics[scale=0.4, trim = 20 20 40 30,clip]{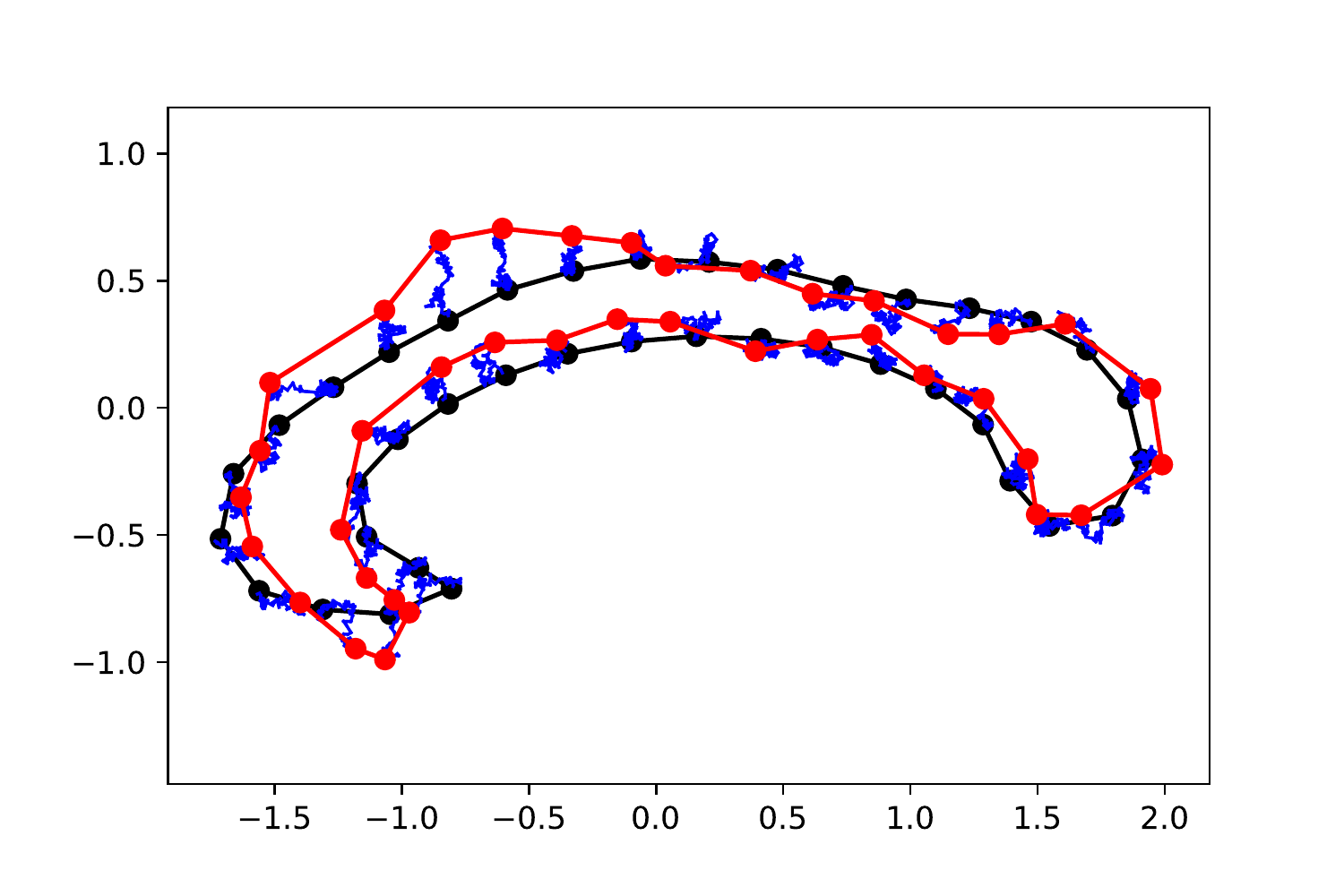}
    \end{minipage}%
    \begin{minipage}{0.5\textwidth}
        \centering
        \includegraphics[scale=0.4,trim = 20 20 40 30,clip]{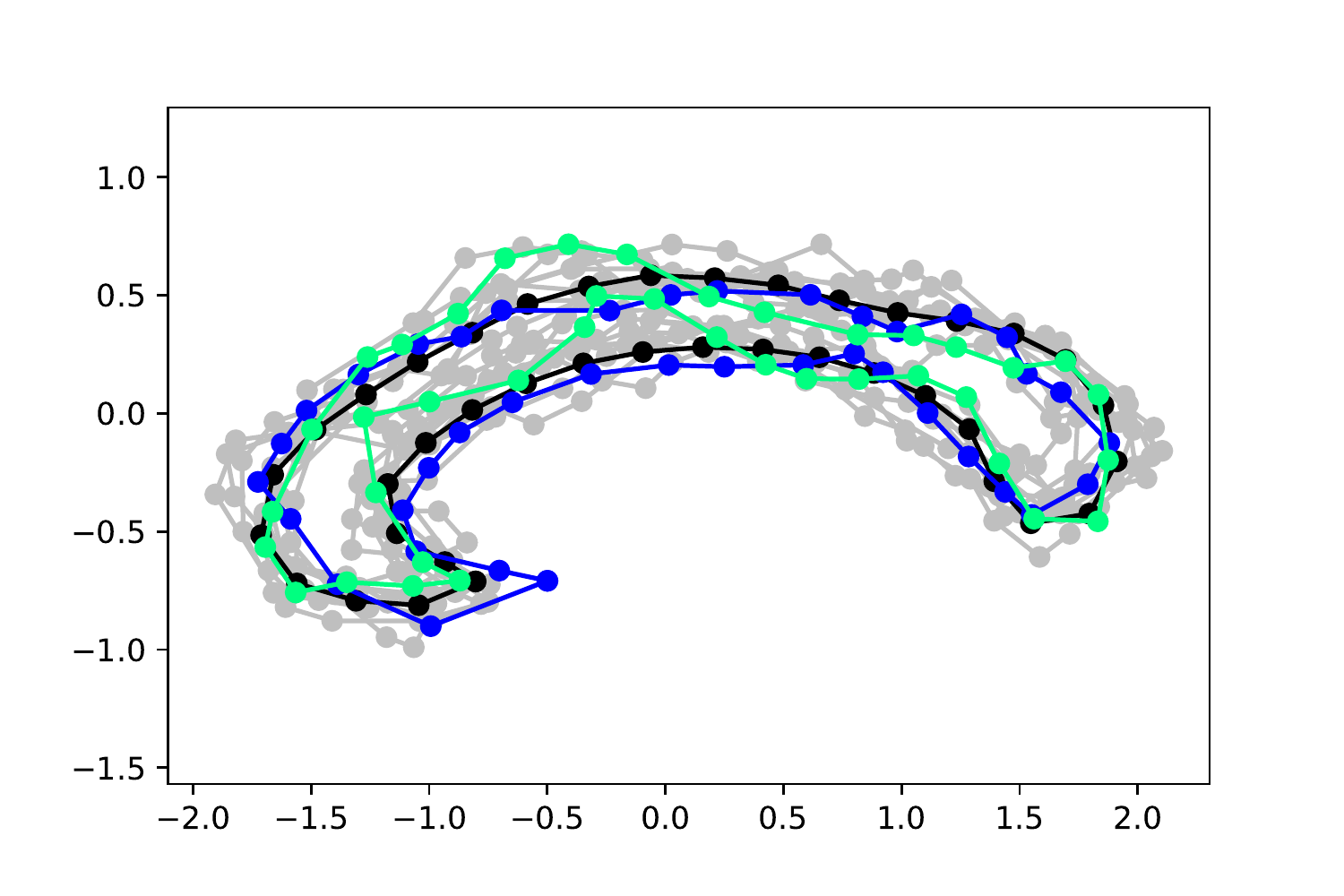}
    \end{minipage}
    \caption{(left) Brownian motion on $\M$. (right) Samples drawn from an isotropic normal distribution defined as the transition distribution of a Brownian motion obtained as a solution to $(\ref{eq:coordB})$.}
    \label{fig:norm}
    \end{center}
\end{figure}

 However, if one is only interested in isotropic Brownian motions, a different definition can be applied. In~\cite{sommer_bridge_2017}, the coordinates of a Brownian motion is defined as solution to the It\^o integral,
\begin{equation}
    dq_t^i = g_q^{kl}\Gamma_{kl}^i dt + \sqrt{g_q^*}^i dW_t.
\label{eq:coordB}
\end{equation}
This stochastic differential equation is implemented in Theano by the following code.
\begin{lstlisting}
"""
Brownian Motion in Coordinates
"""
def @sde_Brownian_coords@(dW,t,q):
    gMsharpq = gMsharp(q)
    X = theano.tensor.slinalg.Cholesky()(gMsharpq)
    det = T.tensordot(gMsharpq,Gamma_gM(q),((0,1),(0,1)))
    sto = T.tensordot(X,dW,(1,0))
    return (det,sto,X)
Brownian_coords = lambda x,dWt: integrate_sde(sde_Brownian_coords,
                                              integrator_ito,x,dWt)
\end{lstlisting}
 An example of an isotropic Brownian motion found by the solution of $(\ref{eq:coordB})$ is shown in Fig. \ref{fig:norm}. 
\vspace{-0.4cm}

In Euclidean statistical theory, a normal distribution can be considered as the transition distribution of a Brownian motion. A similar definition was described in~\cite{sommer_modelling_2015}. Here a normal distribution on $\M$ is defined as the transition distribution of a Brownian motion on $\M$. In Fig. \ref{fig:norm} is shown samples drawn from a normal distribution on $\M$ with mean set to the average CC shown in Fig. \ref{fig:data} and isotropic covariance. The Brownian motions are in this example defined in terms of $(\ref{eq:coordB})$.

\section{Frech\'et Mean on Frame Bundle}
\label{sec:FMeanfm}
\vspace{-0.2cm}
A common task in statistical analysis is to estimate the distribution of data samples. If the observations are assumed to be normally distributed, the goal is to estimate the mean vector and covariance matrix. In~\cite{sommer_modelling_2015}, it was proposed to estimate the mean and covariance of a normal distribution on $\M$ by the Frech\'et mean on the frame bundle. 

Consider Brownian motions on $\M$ defined as the projected stochastic development of Brownian motions on $\R^d$. A normal distribution on $\M$ is given as the transition distribution of a Brownian motion on $\M$.
The initial point for the stochastic development, $u_0 = (q_0,\nu_0)\in F\M$, corresponds to the mean and covariance, i.e. $q_0\in\M$ denotes the mean shape and $\nu_0$ the covariance of the normal distribution. As a consequence, normal distributions with anisotropic covariance can be obtained by letting $\nu_0$ be a non-orthonormal frame. 

In Section \ref{sec:FMean}, the Frech\'et mean on $\M$ was defined as the point $y\in\M$, minimizing the average geodesic distance to the observations. However, as only a sub-Riemannian structure is defined on $F\M$, the logarithm map does not exist and hence the geodesic distance cannot be used to define the Frech\'et mean on $F\M$. Instead, the distance function will be defined based on the most probable paths (MPP) defined in~\cite{sommer_anisotropic}. In this section, a slightly different algorithm for estimating the mean and covariance for a normal distribution is proposed compared to the one defined in~\cite{sommer_modelling_2015}.

Let $u=(q,\nu)\in F\M$ be given such that $q,\nu$ is the mean and covariance of a normal distribution on $\M$. Assume that observations $y_1,\ldots,y_n\in\M$ have been observed and let $p_1,\ldots,p_n\in H^* F\M$. The Frech\'et mean on $F\M$ can then be obtained by optimizing,
\begin{equation*}
    F_{F\M} = \argmin_{(u,p_1,\ldots,p_n)} \frac{1}{n}\sum_{i=1}^n \|p_i\|_{g^*_{FM}}^2 + \frac{\lambda}{n}\sum_{i=1}^n d_{\M}(\pi(\exp_{u}(p_i^\sharp)),y_i)^2 -\frac{1}{2}\log(\det\nu),
\end{equation*}
where $\sharp$ denotes the sharp map on $F\M$ changing a momentum vector in $T^* F\M$ to the corresponding tangent vector in $TF\M$. Notice that the optimization is performed over $u\in F\M$, which is the Frech\'et mean, but also over the momentum vectors $p_1,\ldots,p_n$. When making optimization over these momentum vectors, the geodesics, $\exp_{u}(p_i^\sharp)$, becomes MPPs. The Frech\'et mean on $F\M$ is implemented in Theano and numpy as,

\begin{lstlisting}
"""
Frechet Mean on FM
"""
detg = lambda q,nu: T.nlinalg.Det()(T.tensordot(nu.T,
                 T.tensordot(gM(q),nu,axes=(1,0)),axes=(1,0)))

lossf = lambda q1,q2: 1./d.eval()*np.sum((q1-q2)**2)

def @Frechet_meanFM@(u,p,y0):
     q = u[0:d.eval()]
     nu = u[d.eval():].reshape((d.eval(),rank.eval())) 
     for i in range(n_samples):
         distv[i,:] = lossf(Expfmf(u,p[i,:])[0:d.eval()],y0)
         normp[i,:] = 2*Hfm(u,p[i,:]) # Hamiltonian on FM
     return 1./n_samples*np.sum(normp) 
       +lambda0/n_samples*np.sum(distv**2)-1./2*np.log(detgf(x,u))
\end{lstlisting}

\section{Conclusion}
\label{sec:con}
\vspace{-0.2cm}
In the paper, it has been shown how different concepts in differential geometry and non-linear statistics can be implemented using the Theano framework. Integration of geodesics, computation of Christoffel symbols and parallel transport, stochastic development and Fre\'chet mean estimation were considered and demonstrated on landmark shape manifolds. In addition, we showed how the Fre\'chet mean on the frame bundle $F\M$ can be computed for estimating the mean and covariance of an anisotropic normal distribution on $\M$. 

Theano has, for the cases shown in this paper, been a very efficient framework for implementation of differential geometry concepts and for non-linear statistics in a simple and concise way yet allowing efficient and fast computations. We emphasize that Theano is able to perform calculations on high-dimensional manifolds using either CPU or GPU computations. In the future, we plan to extend the presented ideas to derive Theano implementations of differential geometry concepts in more general fiber bundle and Lie group geometries.

\vspace{0.1cm}

\textbf{Acknowledgements.} This work was supported by Centre for Stochastic Geometry and Advanced Bioimaging (CSGB) funded by a grant from the Villum foundation.

\bibliographystyle{plain}
%\bibliography{Line}

\end{document}